%% file: Paper_50.tex
\documentclass[runningheads]{llncs}
\usepackage{graphicx}
\usepackage{amssymb}
\usepackage{amsmath}
\usepackage{microtype}
\usepackage{booktabs}
\usepackage{multirow}
\usepackage{array}
\usepackage{hyperref}
\usepackage{cite}
\usepackage{microtype}
\usepackage{xcolor}
\definecolor{mycolor}{HTML}{FF6600}
\definecolor{mycolor2}{HTML}{6699CC}
\definecolor{mycolor3}{HTML}{CC0000}

\begin{document}
\title{StarGAN-ZSVC: Towards Zero-Shot Voice Conversion in Low-Resource Contexts\thanks{This work is supported in part by the National Research Foundation of South Africa (grant number: 120409) and a Google Faculty Award for HK.}}
\titlerunning{StarGAN-ZSVC}
\author{Matthew Baas\orcidID{0000-0003-3001-6292} \and
Herman Kamper\orcidID{0000-0003-2980-3475}}
\authorrunning{M. Baas and H. Kamper}

\institute{E\&E Engineering, Stellenbosch University, Stellenbosch, South Africa\\
\email{\{20786379,kamperh\}@sun.ac.za}}

\maketitle  

\begin{abstract}
Voice conversion is the task of converting a spoken utterance from a source speaker so that it appears to be said by a different target speaker while retaining the linguistic content of the utterance. Recent advances have led to major improvements in the quality of voice conversion systems. 
However, to be useful in a wider range of contexts, voice conversion systems would need to be (i)~trainable without access to parallel data, (ii) work in a zero-shot setting where both the source and target speakers are unseen during training, and (iii)~run in real time or faster. Recent techniques fulfil one or two of these requirements, but not all three. This paper extends recent voice conversion models based on generative adversarial networks (GANs), to satisfy all three of these conditions. We specifically extend the recent StarGAN-VC model by conditioning it on a speaker embedding (from a potentially unseen speaker). This allows the model to be used in a zero-shot setting, and we therefore call it StarGAN-ZSVC. We compare StarGAN-ZSVC against other voice conversion techniques in a low-resource setting using a small 9-minute training set.
Compared to AutoVC---another recent neural zero-shot approach---we observe that StarGAN-ZSVC gives small improvements in the zero-shot setting, showing that real-time zero-shot voice conversion is possible even for a model trained on very little data. Further work is required to see whether scaling up StarGAN-ZSVC will also improve zero-shot voice conversion quality in high-resource contexts.

\end{abstract}


\input{1_intro}

\input{2_related_work}

\input{3_model}

\input{4_experiments}

\input{5_results}

\input{6_conclusion}

%
%
%

\bibliography{References}{}
\bibliographystyle{splncs04}
\end{document}

%% file: 1_intro.tex
\section{Introduction}
Voice conversion is a speech processing task where speech from a source speaker is transformed so that it appears to come from a different target speaker while preserving linguistic content.
A fast, human-level voice conversion
system has significant applications across several industries, from those in privacy and identity protection \cite{privacy} to those of voice mimicry and disguise \cite{vc_vc_spk_id,disguise}.
It can also be essential for addressing downstream speech processing problems in low-resource contexts where training data is limited:
it could be used to augment training data by converting the available utterances to novel speakers---effectively increasing the diversity of training data and improving the quality of the resulting systems.

Recent techniques have improved the quality of voice conversion significantly, in part due to the Voice Conversion Challenge (VCC) and its efforts to concentrate disparate research efforts \cite{vcc2020}. 
Some techniques are beginning to achieve near human-level quality in conversion outputs.
However, much of the advances and improvements in quality are limited in their practical usefulness because they fail to satisfy several requirements that would be necessary for practical use, particularly in low-resource settings.

First, a practical voice conversion system should be trainable on non-parallel data.
That is, training data should not need to contain utterances from multiple speakers saying the same words -- such a setting is known as a parallel data setting.
Non-parallel data is the converse, where the different utterances used to train the model do not contain the same spoken words.
Parallel data is difficult to collect in general, and even more so for low-resource language (those which have limited digitally stored corpora). 
Second, a practical system should be able to convert speech to and from speakers which have not been seen during training.
This is called \textit{zero-shot} voice conversion.
Without this requirement, a  system would need to be retrained whenever speech from a new speaker is desired.
Finally, for a number of practical applications, a voice conversion system needs to run at least in real-time. 
For data augmentation in particular, having the system run as fast as possible is essential for it to be practical in the training of a downstream speech model.

With these requirements in mind, we look to extend existing state-of-the-art voice conversion techniques.
We specifically extend the recent StarGAN-VC2 \cite{stargan-vc2} approach to the zero-shot setting, proposing the new StarGAN-ZSVC model.
StarGAN-ZSVC achieves zero-shot prediction by using a speaker encoding network to generate speaker embeddings for potentially unseen speakers; these embeddings are then used to condition the model at inference time.
  
Through objective and human evaluations, we show that StarGAN-ZSVC performs better than simple baseline models and similar or better than the recent AutoVC zero-shot voice conversion approach~\cite{autovc} across a range of evaluation metrics.
More specifically, it gives similar or better performance in all zero-shot settings considered, and does so more than five times faster than AutoVC.

%% file: 2_related_work.tex
\section{Related Work}
A typical voice conversion system operates in the frequency domain, first converting an input utterance into a spectrogram and then using some model to map the spectrogram spoken by a source speaker to that of one spoken by a target speaker.
The output spectrogram is then converted to a waveform in the time-domain using
a vocoder \cite{metaanalysis}.
In this paper, we denote spectrogram sequences as $X = [\mathbf{x}_1, \mathbf{x}_2, ..., \mathbf{x}_T]$, where the spectrogram contains $T$ frames, and each frame $\mathbf{x}_i$ consists of some number of frequency bins.
In our case, we use 80-dimensional Mel-scale frequency bins, i.e.\  
$\mathbf{x}_i \in \mathbb{R}^{80}$.

Some models~\cite{vtln-vc,stargan-vc}
use parametric algorithms like the WORLD vocoder \cite{WORLD-vocoder} to convert the output spectrogram 
back to a time-domain waveform.
Others use neural vocoders, which can be 
divided into autoregressive models, such as those of the WaveNet family \cite{wavenet}, and non-autoregressive models, such as MelGAN \cite{MelGAN}. 

Voice conversion models themselves can be classified on several levels.
Older techniques rely on rule-based techniques \cite{rules-vc-1,rules-vc-2} while newer models rely on statistical techniques and often make extensive use of deep neural networks \cite{metaanalysis}.
Models can also be classified into traditional models that can only perform one-to-one voice conversion, such as the recurrent DBLSTM-RNN \cite{BDLSTM} and Gaussian mixture based models \cite{gmm-vc-1,gmm-vc-2}, to newer models like those using variational autoencoders \cite{autovc,autoencoder-vc2} and vector quantized neural networks \cite{vqvae-comparisons} to allow for many-to-many conversion where a single model can convert between several possible source-target speaker pairings.

Finally, the recent AutoVC model~\cite{autovc} (Section~\ref{sec:autovc}) emerged as the first model to be able to perform zero-shot voice conversion where either the source or target speaker is unseen during training.
For our new model, we also take inspiration from recent work on speaker encoding networks trained for speaker verification~\cite{GE2E}, as well as the StarGAN-VC and -VC2 \cite{stargan-vc,stargan-vc2} models (Section~\ref{sec:stargan}). Concretely, we attempt to combine these in a new zero-shot voice conversion~model.

\subsection{StarGAN and Voice Conversion}
\label{sec:stargan}

Generative adversarial networks (GANs) train two separate networks: a generator and a discriminator.
The generator is trained to produce realistic outputs (i.e.\ it should aim to accurately approximate some function), while the discriminator is trained to discern true outputs from ones produced by the generator.
Part of the generator's objective is to fool the discriminator and to essentially maximize its loss metric, while the discriminator is trained to do the opposite.

One set of successful voice conversion techniques relies on re-purposing the StarGAN image-to-image translation technique \cite{choi2018stargan} for voice conversion.
In particular, StarGAN-VC2 \cite{stargan-vc2} extends upon StarGAN-VC \cite{stargan-vc} by training a single generator model to perform many-to-many voice conversion using speaker-dependent modulation factors in so-called conditional instance normalization layers \cite{CIN}.
The model's
generator $G(X_\text{src}, \mathbf{s}_\text{src}, \mathbf{s}_\text{trg})$ converts a spectrogram $X_{\text{src}}$ from a source speaker to a target speaker, producing the converted output $X_{\text{src}\rightarrow \text{trg}}$.
The source and target speaker identities are given as one-hot vectors,  $\mathbf{s}_\text{src}$ and $\mathbf{s}_\text{trg}$, respectively.
The model's discriminator $D(X, \mathbf{s}_\text{src}, \mathbf{s}_\text{trg})$ takes an input spectrogram $X$ and returns a scalar. 
Intuitively, the generator is trained to force the discriminator's output when given converted spectrograms to be high, while the discriminator is trained to make its output low when given converted outputs and high when given original spectrograms.

More formally, the generator $G$ is trained to minimize the loss $\mathcal{L} = \lambda_{\text{id}}\mathcal{L}_{\text{id}} + \lambda_{\text{cyc}}\mathcal{L}_{\text{cyc}} + \mathcal{L}_{G-\text{adv}}$.
The first term, $\mathcal{L}_{\text{id}}$ is an identity loss term.
It aims to minimize the difference between the input and output spectrogram when the model is made to keep the same speaker identity, i.e.\ convert from speaker A to speaker~A. 
It is defined by the $L_2$ loss:
\begin{equation}
    \mathcal{L}_{\text{id}} = ||G(X_\text{src}, \mathbf{s}_{\text{src}}, \mathbf{s}_{\text{src}}) - X_\text{src}||_2
\end{equation}
Next, many-to-many voice conversion systems like StarGAN-VC2 can perform \textit{cyclic mappings}, whereby the model is made to convert an input utterance from a source speaker to a target speaker, and then convert the output utterance back to the source speaker.
The second term of the loss aims to minimize the cyclic reconstruction error between the cyclic mapping and original spectrogram \cite{choi2018stargan}:
\begin{equation}
    \mathcal{L}_{\text{cyc}} = ||X_\text{src} - G(G(X_\text{src}, \mathbf{s}_{\text{src}}, \mathbf{s}_{\text{trg}}), \mathbf{s}_{\text{trg}}, \mathbf{s}_{\text{src}})||_1
\end{equation}
Finally, the adversarial loss term $\mathcal{L}_{G-\text{adv}}$ is added based on the LSGAN \cite{LSGAN} loss.
It defines two constants $a$ and $b$, whereby $G$'s loss tries to push $D$'s output for converted utterances closer to $a$, while $D$'s loss function tries to push $D$'s output for converted utterances closer to $b$ and its output for real outputs closer to $a$. 
Concretely, $G$'s adversarial loss is defined as
\begin{equation}
    \mathcal{L}_{G-\text{adv}} = \left( D(G(X_\text{src}, \mathbf{s}_{\text{src}},\mathbf{s}_{\text{trg}}), \mathbf{s}_{\text{src}}, \mathbf{s}_{\text{trg}}) - a \right)^2.
\end{equation}
while, the discriminator $D$ is trained to minimize the corresponding LSGAN discriminator loss:
\begin{equation}
    \mathcal{L}_{D-\text{adv}} = \left( D(G(X_\text{src}, \mathbf{s}_{\text{src}},\mathbf{s}_{\text{trg}}), \mathbf{s}_{\text{src}}, \mathbf{s}_{\text{trg}}) - b \right)^2 + \left( D(X_\text{src}, \mathbf{s}_{\text{trg}}, \mathbf{s}_{\text{src}}) - a \right)^2
\end{equation}
In~\cite{stargan-vc2}, the authors set the scalar coefficients to be $\lambda_\text{id}=5$, and $\lambda_\text{cyc}=10$.
The original study~\cite{stargan-vc2} does not mention how $a$ and $b$ are set (despite these greatly affecting training); we treat them as hyperparameters.
Note that the true target spectrogram $X_\text{trg}$ does not appear in any of the equations -- this is what allows StarGAN-VC2 to be trained with non-parallel data where the source utterance $X_\text{src}$ has no corresponding utterance from the target speaker.

StarGAN-VC2 uses a specially designed 2-1-2D convolutional architecture for the generator, as well as a projection discriminator \cite{projection_discriminator} which comprises of a convolutional network (to extract features) followed by an inner product with an embedding corresponding to the source/target speaker pair.
For the generator, a new form of modulation-based conditional instance normalization was introduced in \cite{stargan-vc2}. This allows the speaker identity (which is provided as a one-hot vector) to multiplicatively condition the channels of an input feature.
According to \cite{stargan-vc2}, this special layer is a key component in achieving high performance in StarGAN-VC2.

We use these building blocks for our new zero-shot approach. Concretely, the one-hot speaker vectors in StarGAN-VC2 are replaced with continuous embedding vectors obtained from a separate speaker encoding network (which can be applied to arbitrary speakers), as outlined in Section~\ref{sec:starganzsvc}.

\subsection{AutoVC}
\label{sec:autovc}
Zero-shot voice conversion was first introduced in 2019 with the
AutoVC model~\cite{autovc}, which remains one of only a handful of models that can perform zero-shot conversion (see e.g.\ \cite{convoice} for a very recent other example).
For AutoVC, zero-shot conversion is achieved by using an autoencoder with a specially designed bottleneck layer which forces the network's encoder to only retain linguistic content in its encoded latent representation.
The model then uses a separate recurrent speaker encoder model $E(X)$, originally proposed for speaker identification \cite{GE2E}, to extract a speaker embedding $\mathbf{s}$ from an input spectrogram.
These speaker embeddings are then used to supply the missing speaker identity information to the decoder which, together with the linguistic content from the encoder, allows the decoder to synthesize an output spectrogram for an unseen speaker. 

Formally, the full encoder-decoder model is trained to primarily minimize two terms.
The first term is an $L_2$ reconstruction loss between the decoder output spectrogram $X_{\text{src}\rightarrow\text{src}}$ and input spectrogram $X_{\text{src}}$, with the source speaker's encoding (from the speaker encoder) provided to both the encoder and decoder.
The second term is an $L_1$ loss between the speaker embedding of the decoder output $E(X_{\text{src}\rightarrow\text{src}})$ and the original speaker embedding $\mathbf{s}_\text{src} = E(X_{\text{src}})$.
The encoder and decoder consists of convolutional and Long Short-Term Memory (LSTM) \cite{lstm} recurrent layers which are carefully designed
to ensure that no speaker identity information is present in the encoder output.

As with StarGAN-VC and StarGAN-VC2, a corresponding parallel target utterance $X_\text{trg}$ does not appear in any of the loss terms, allowing AutoVC to be trained without parallel data. 
Zero-shot inference is performed by using the speaker encoder to obtain embeddings for new utterances from unseen speakers, which is then provided to the decoder instead of the embedding corresponding to the source speaker, causing the decoder to return a converted output. 
We use this same idea of using an encoding network to obtain embeddings for unseen speakers in our new GAN-based approach, which we describe next.

%% file: 3_model.tex
\section{StarGAN-ZSVC}
\label{sec:starganzsvc}

While StarGAN-VC and StarGAN-VC2 allows training with non-parallel data and runs sufficiently fast, it is limited in its ability to only perform voice conversion for speakers seen during training: while parallel $X_{\text{src}}$ and $X_{\text{trg}}$ utterance pairs are not required, the model can only synthesize speech for target speaker identities (specified as one-hot vectors) seen during training.
This could preclude the use of these models in many practical situations where zero-shot conversion is required between unseen speakers.
Conversely, AutoVC allows for such zero-shot prediction and is trained on non-parallel data, but it is implemented with a slow vocoder and
its performance suffers when trained on very little data. Combining the strengths of both of these methods, we propose the \textit{StarGAN zero-shot voice conversion} model -- StarGAN-ZSVC.

\subsection{Overcoming the Zero-shot Barrier}
To achieve voice conversion between multiple speakers, the original StarGAN-VC2 model creates an explicit embedding vector for each source-target speaker pairing, which is incorporated
as part of the generator $G$ and discriminator $D$.
This requires that each source-target speaker pairing is seen during training so that the corresponding embedding exists and has been trained -- prohibiting zero-shot voice conversion.
To overcome this hurdle, we instead infer separate source and target speaker embeddings, $\mathbf{s}_{\text{src}}$ and $\mathbf{s}_{\text{trg}}$, using a speaker encoder network $E$ -- similar to the approach followed in AutoVC (Section~\ref{sec:autovc}).

\begin{figure}[!t]
\includegraphics[width=\textwidth]{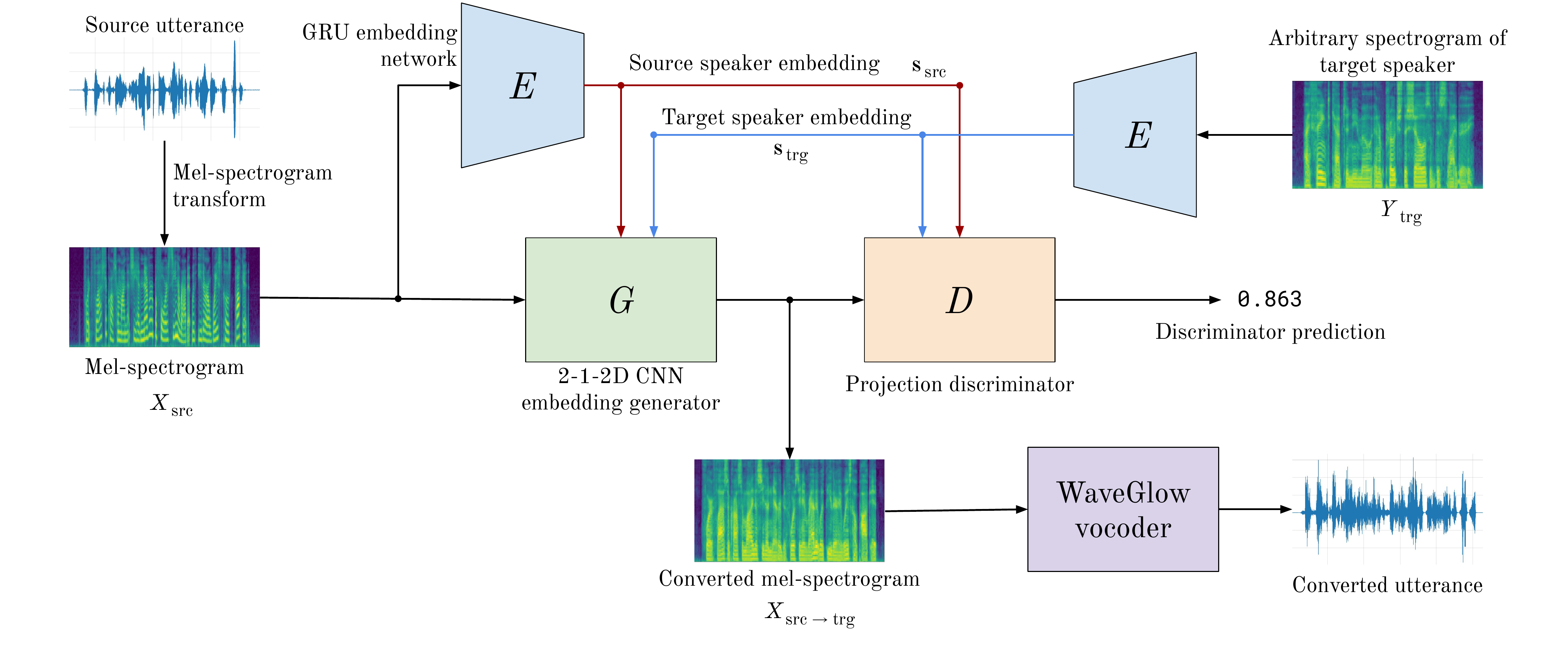}
\caption{The StarGAN-ZSVC system framework. The speaker encoder network $E$ and the WaveGlow vocoder are pretrained on large speech corpora, while the generator $G$ and discriminator $D$ are trained on a 9-minute subset of the VCC dataset. During inference, arbitrary utterances for the source and target speaker are used to obtain source and target speaker embeddings, $\mathbf{s}_{\textrm{src}}$ and $\mathbf{s}_{\textrm{trg}}$. 
} \label{fig:system-diagram}
\end{figure}

This framework is shown in Figure~\ref{fig:system-diagram}.
Utterances from unseen speakers (i.e.\  $X_{\text{src}}$ and $Y_{\text{trg}}$) are fed to the speaker encoder $E$, yielding embeddings for these new speakers, which are then used to condition $G$ and $D$, 
thereby enabling zero-shot conversion. 
The generator uses these embeddings to produce a converted Mel-spectrogram $X_{\text{src}\rightarrow \text{trg}}$ from a given source utterance's Mel-spectrogram $X_{\text{src}}$.

$E$ is trained on a large corpus 
using a GE2E loss \cite{GE2E} which aims to simultaneously maximize distances between embeddings from different speakers while minimizing the distances between embeddings from utterances of the same speaker.
NVIDIA's WaveGlow \cite{waveglow} is used, which does not require any speaker information for vocoding and thus readily allows zero-shot conversion.

\subsection{Overcoming the Speed Barrier}
The speed of the full voice conversion system during inference is bounded by 
(a)~the speed of the generator $G$; 
(b)~the speed of converting the utterance between time and frequency domains, consisting of the initial conversion from time-domain waveform to Mel-spectrogram and the speed of the vocoder; 
and (c)~the speed of the speaker encoder $E$.
To ensure that the speed of the full system is at least real-time, each subsystem needs to be
faster than real-time.

\subsubsection{(a) Generator Speed.}
For the generator $G$ to be sufficiently fast, we design it to only include convolution, linear, normalization, and upscaling layers as opposed to a recurrent architecture like those used in AutoVC \cite{autovc}. 
By ensuring that the majority of layers are convolutions, we obtain better-than real-time speeds for the generator.

\subsubsection{(b) Vocoder and Mel-spectrogram Speed.} 
The choice of vocoder greatly affects computational cost. 
Higher-quality methods, such as those from the WaveNet family~\cite{wavenet}, are typically much slower than real-time, while purely convolutional methods such as MelGAN~\cite{MelGAN} are much faster 
but has poorer quality.
Often the main difference between the slower and faster methods is again the presence of traditional recurrent layers in the vocoder architecture. 

We opt for a reasonable middle-ground choice with the WaveGlow vocoder, which does have recurrent connections but does not use any recurrent layers with dense multiplications such as LSTM or Gated Recurrent Unit (GRU, another kind of recurrent cell architecture \cite{gru}) layers. We specifically use a pretrained WaveGlow network, as provided with 
the original paper \cite{waveglow}. 
Furthermore, the speed of the Mel-spectrogram transformation for the input audio is well faster than real-time due to the efficient nature of the fast Fourier transform and the multiplication by Mel-basis filters.

\subsubsection{(c) Speaker Encoder Speed.} 
The majority of research efforts into obtaining speaker embeddings involve models using slower recurrent layers, often making these encoder networks the bottleneck. 
We also make use of a recurrent stacked-GRU network as our speaker embedding network $E$.
However, we only need to obtain a single speaker embedding to perform 
any number of conversions involving that speaker.
We therefore treat this as a preprocessing step where we apply $E$ to a few arbitrary utterances from the target and source speakers, averaging the results to obtain target and source speaker embeddings, and use those same embeddings for all subsequent conversions.

We also design the speaker embeddings to be 256-dimensional vectors of unit length. If we were to use StarGAN-ZSVC downstream for data augmentation (where we want speech from novel speakers), we could then simply sample random unit-length vectors of this dimensionality  to use with the generator.

\begin{figure}[!t]
\includegraphics[width=\textwidth]{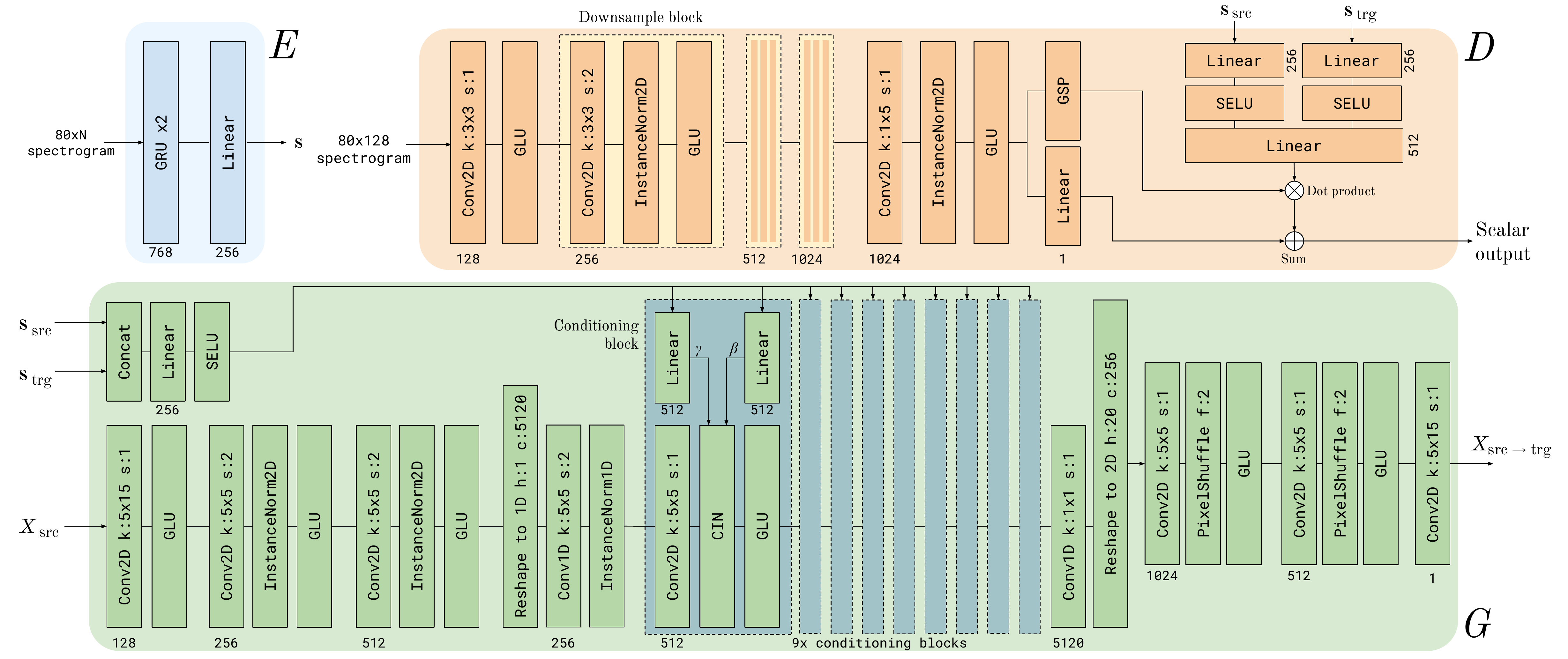}
\caption{StarGAN-ZSVC's network architectures. The speaker encoder $E$ is a recurrent network similar to that used in the original GE2E paper, while the generator $G$ and discriminator $D$ are modified versions from the original StarGAN-VC2 architecture. Within layers, \texttt{k} and \texttt{s} represent kernel size and stride (for convolutions), \texttt{f} is the scaling factor (for pixel shuffle), and \texttt{h} and \texttt{c} are the height and channels of the output (for reshape layers). A number alongside a layer indicates the number of output channels (for convolutions), or output units (for linear and GRU layers). GLU layers split the input tensor in half along the \textit{channels} dimension. GSP, GLU, and SELU indicate global sum pooling, gated linear units, and scaled exponential linear units, respectively.} \label{fig:architecture}
\end{figure}

\subsection{Architecture}
With the previous considerations in mind, we design the generator $G$, discriminator $D$, and encoder network $E$, as shown in Figure~\ref{fig:architecture}.
The generator and discriminator are adapted from StarGAN-VC2 \cite{stargan-vc2}, while the speaker encoder
is adapted from the original model proposed for speaker identification \cite{GE2E}. 
Specifically, for $E$ 
we use a simple stacked GRU model, while for $D$ we use a projection discriminator \cite{projection_discriminator}. 
For $G$, we use the 2-1-2D generator from StarGAN-VC2 with a modified central set of layers, denoted by the \textit{Conditional Block} in the figure. 

These conditional blocks are intended to provide the network with a way to modulate the channels of an input spectrogram, with modulation factors conditioned on the specific source and target speaker pairing.
They utilize a convolutional layer 
followed by a modified conditional instance normalization layer \cite{CIN} and a gated linear unit \cite{glu}.

The modified conditional instance normalization layer performs the following operation on an input feature vector $\mathbf{f}$:
\begin{equation}
\text{CIN}(\mathbf{f}, \gamma, \beta) 
= \gamma \left( \frac{\mathbf{f} - \mu(\mathbf{f})}{\sigma (\mathbf{f}) } \right) + \beta
\end{equation}
where $\mu(\mathbf{f})$ and $\sigma(\mathbf{f})$ are respectively the scalar mean and standard deviation of vector $\mathbf{f}$, while $\gamma$ and $\beta$ are computed using two linear layers which derive their inputs from the speaker embeddings, as depicted in Figure~\ref{fig:architecture}. The above
is computed separately for each channel when the input feature contains multiple~channels.

For the discriminator, the source and target speaker embeddings are also fed through several linear layers and activation functions to multiply with the pooled output of $D$'s main branch.

%% file: 4_experiments.tex
\section{Experimental Setup}

We compare StarGAN-ZSVC to other voice conversion models using the voice conversion challenge (VCC) 2018 dataset \cite{vcc2018}, which contains parallel recordings of native English speakers from the United States. 
Importantly, we \textit{do not} train StarGAN-ZSVC or the AutoVC model (to which we compare) using parallel input-output examples.
However, the traditional baseline models (below) do require parallel data.
All training and speed measurements are performed on a 
single {NVIDIA} RTX 2070 SUPER GPU. 

\subsection{Dataset}
\label{sec:dataset}
The VCC 2018 dataset was recorded from 8 speakers, each speaking 81 utterances from the same transcript. 4 speakers are used for training and 4 for testing.
To emulate a low-resource setting, we use a 9-minute subset of the VCC 2018 training dataset for StarGAN-ZSVC and AutoVC.
This corresponds to 90\% of the utterances from two female (F) and two male (M) speakers (VCC2SF1, VCC2SF2, VCC2SM1, and VCC2SM2).
This setup is in line with existing evaluations on VCC 2018 \cite{stargan-vc2}, allows for all combinations of inter- and intra-gender conversions, and allows for zero-shot evaluation on the 4 remaining unseen speakers.

In contrast to StarGAN-ZSVC and AutoVC, some of the baseline models only allow for one-to-one conversions, i.e. they are trained on parallel data and can only convert from seen speaker A to seen speaker B. 
We therefore train the baseline models on a single source-target speaker mapping (from VCC2SF1 to VCC2SM2), using 90\% of the parallel training utterances for this speaker pair.

All utterances are resampled to 22.05~kHz and then converted to log Mel-spectrograms with a window and hop length of 1024 and 256 samples, respectively. 
During training, for each batch we randomly sample a $k$-frame sequence from each spectrogram, where $k$ is randomly sampled from multiples of 32 between 96 to 320 (inclusive).
This is done for all models to make it robust to utterance length, with the exception of StarGAN-ZSVC, which requires fixed-size input for its discriminator.
This leads to slightly worse performance for
StarGAN-ZSVC on long or silence-padded sequences.
For a fair comparison, we therefore only consider non-silent frames of the target utterance.

\subsection{Speaker Encoder and Vocoder Setup}
The same WaveGlow vocoder is used
to produce output waveforms
for all networks
to ensure a fair comparison. 
The WaveGlow model is pretrained on a large external corpus, as provided by the original paper \cite{waveglow}. 
However, since all models use log Mel-spectrogram inputs and produce log Mel-spectrogram outputs, we rather perform all quantitative comparisons on the spectrograms of each utterance (instead of waveforms), in order to minimize the vocoder's confounding effect. 

Our WaveGlow implementation takes approximately 240~ms to produce one second of vocoded audio (taking a spectrogram as input).
For the full voice conversion system to be faster than real-time, this means that the combined inference speed of the remaining sub-networks needs to be well under 700ms/s, or preferably significantly faster if used for data augmentation.

The speaker encoder is trained on 90\% of the utterances from a combined set consisting of the VCTK \cite{vctk}, VCC 2018 \cite{vcc2018}, LibriSpeech \cite{librispeech}, and the English CommonVoice 2020-06 datasets.\footnote{Available under a CC-0 license at \url{https://commonvoice.mozilla.org/en/datasets}.} 
It is trained with the Adam optimizer \cite{adam} for 8 epochs with 8 speakers per batch, and 6 utterances per speaker in each batch.
We start with a learning rate of $4\times10^{-4}$ and adjust it down to $3\times10^{-7}$ in the final epoch.
Embeddings for speakers are precomputed by taking the average embedding over 4 arbitrary utterances for each speaker.

\subsection{Baseline Models}
We train 4 baseline models for comparison, all using the Adam optimizer.
The first three are traditional one-to-one conversion models, consisting of a simple linear model, a DBLSTM model \cite{BDLSTM}, and a UNet model \cite{unet}. 
The final model that we compare to is the AutoVC model, 
which is able to do zero-shot many-to-many conversion (Section~\ref{sec:autovc}). 
We compare AutoVC to StarGAN-ZSVC on all variants of seen/unseen source/target pairings. 
Each network is trained according to the method developed by
Smith \cite{smith:cyclic} by evaluating the decrease in loss every few hundred epochs for different learning rates, and updating the learning rate to correspond to the largest decrease in loss. 
This process is repeated until the validation loss begins to increase, after which training is terminated.

All one-to-one models are trained in the same way, taking the source spectrogram $X_{\text{src}}$ as input and trained with an $L_1$ loss (which we found to produce better results than the $L_2$ loss) to predict the ground-truth target spectrogram $X_{\text{trg}}$. 
The linear model consists of 4 convolutional layers with output channels and kernel sizes of (200, 5), (200, 5), (100, 3), and (1, 3), respectively. 
The DBLSTM model is based on the original 
paper \cite{BDLSTM}, but we do not use any time-alignment techniques (such as dynamic time warping). 
The model consists of 4 stacked bidirectional LSTM layers with a hidden size of 256 and a dropout of 0.3, followed by a final linear projection layer to bring the output dimension back to 80. The network is trained with a batch size of 8.
The UNet model is built based on the structure of XResNet \cite{xresnet} using the method defined in the Fast.ai library \cite{fastai:unet}.

Finally, AutoVC is trained using the same loss function as in the original paper \cite{autovc}. It is trained with a batch size of 4 for 4700 epochs. 
The speaker encoder used is the pretrained encoder $E$ described above. 

\subsection{StarGAN-ZSVC Training}
We train StarGAN-ZSVC using the same Adam optimizer and learning rate scheduling technique of Smith \cite{smith:cyclic}.
Furthermore, we employ several tricks for successfully training GANs: 
(i) gradients in $G$ and $D$ are clipped to have a maximum norm of 1; 
(ii) the discriminator's learning rate is made to always be half of the generators learning rate; 
(iii) the number of iterations training the discriminator versus generator is updated every several hundred epochs to ensure that the discriminator's loss is always roughly a factor of 10 lower than the adversarial term of the generator's loss; and
(iv) dropout with a probability of 0.3 is added to the input of $D$ after the first 3000 epochs (if added earlier it causes artifacts and destabilizes training). 

The loss function used is the same as that of StarGAN-VC2, with the exception that the term $\mathcal{L}_\text{cyc}$ (see Section~\ref{sec:stargan}) is squared in our model, which we found to give superior results. 
We set $a = 1$, and $b=0$ for the LSGAN constants, and $\lambda_{\text{cyc}} = 10$, $\lambda_{\text{id}} = 5$ for loss coefficients, being adjusted downwards during training in the same manner as in \cite{stargan-vc2}. 

\subsection{Evaluation}

We compare converted output spectrograms to their ground-truth target spectrograms on the test dataset using several objective metrics.
To account for different speaking rates, we first use dynamic time warping (DTW) to align the converted spectrogram to the target spectrogram, and then perform comparisons over non-silent regions of the \textit{target} utterance. Non-silent regions are defined as those 80-dimensional spectrogram frames where the mean vector element value is greater than -10dB.

In addition to computing the mean absolute error (MAE) and mean square error (MSE) between spectrograms, we also compute a \textit{cosine similarity} by finding the cosine distance between each 80-dimensional source/target frame pair of the Mel-spectrograms and then computing the mean cosine distance over all non-silent frames (after DTW alignment). 
This metric, denoted as $\cos(\theta)$, gives an additional measure of conversion quality. 

Finally, to quantitatively measure speaker similarity (i.e.\ determining whether the generated spectrogram sounds like the target speaker), we define a new metric using the speaker encoder.
We compute speaker embeddings for the target and output converted spectrograms using the speaker encoder $E$. 
The norm of the difference between these vectors, $||E(X_{\text{src}\rightarrow \text{trg}}) - E(X_{\text{trg}})|| = e_{\text{norm}}$, is then used as a measure of speaker similarity, with a smaller norm corresponding to greater similarity between the converted and target spectrogram.

We also perform a subjective listening test with 12 proficient English speakers to assess how well StarGAN-ZSVC compares to AutoVC across various zero-shot settings.
Each participant rated the naturalness of 144 utterances from 1 (bad) to 5 (excellent) where the utterance order and naming is randomized.
The 144 utterances consist of 8 converted utterances for each seen/unseen source/target speaker pairing, for both AutoVC and StarGAN-ZSVC.
A further 14 utterances are included to find a baseline rating for raw and vocoded audio.
The ratings for each subset are averaged to find a mean opinion score (MOS), which serves as a measure of conversion quality.

%% file: 5_results.tex
\section{Experiments}
We perform two sets of experiments.
First we perform an evaluation on seen speakers, where we compare StarGAN-ZSVC to all other models to obtain an indication of both speed and performance.
We then compare StarGAN-ZSVC with AutoVC for zero-shot voice conversion, 
looking at both the output and cyclic reconstruction error.
We encourage the reader to listen to the demo samples\footnote{\url{https://rf5.github.io/sacair2020/}} for the zero-shot models. 

\subsection{Seen-to-seen Conversion} \label{sec:experiments_seen}

In the first set of comparisons, we evaluate
performance for test utterances where other utterances from both the source and target speaker have been seen during training. I.e., while the models have not been trained on these exact test utterances, they have seen the speakers during training.
There is, however, a problem in directly comparing the one-to-one models (traditional baselines) to the many-to-many models (AutoVC and StarGAN-ZSVC).
The one-to-one models are trained on parallel data, always taking in utterances from one speaker as input (VCC2SF1 in our case) and always producing output from a different target speaker (VCCSM2).

\begin{table}[!b]
    \centering
    \renewcommand{\arraystretch}{1.2}
    \caption{
    Objective evaluation results when converting between speakers where both the source and target speaker are seen during training. 
    For all metrics aside from cosine similarity, lower is better.
    Speed is measured as the time (in milliseconds) required to convert one second of input audio.
    The first StarGAN-ZSVC and AutoVC entries correspond to evaluations on the one-to-one test utterances, while the final two starred entries correspond to metrics computed when using test utterances from all seen training speakers for the many-to-many models.}
    \label{tab:seen-to-seen-results}
    \begin{tabular}{l@{\hspace{0.3cm}}c@{\hspace{0.3cm}}c@{\hspace{0.3cm}}c@{\hspace{0.3cm}}c@{\hspace{0.3cm}}c}
    
    \toprule
    Model & MAE & MSE & $\cos(\theta)$ & $e_{\text{norm}}$ & Speed (ms/s)\\
    \midrule
    Linear & 1.277 & 2.689 & 0.980 & 0.860 & \textbf{0.15}\\
    DBLSTM & 1.329 & 3.102 & 0.982 & 0.496 & 12.52\\
    UNet & 1.370 & 3.347 & 0.980 & 0.545 & 100.5 \\
    AutoVC & \textbf{0.993} & \textbf{1.756} & \textbf{0.987} & \textbf{0.259} & 10.99\\
    StarGAN-ZSVC & 1.092 & 2.101 & 0.977 & 0.513 & 1.88 \\
    \midrule
    AutoVC* & \textbf{1.000} & \textbf{1.783} & \textbf{0.987} & \textbf{0.321} & 10.99\\
    StarGAN-ZSVC* & 1.008 & 1.863 & 0.983 & \textbf{0.321} & 1.88 \\
    \bottomrule
    \end{tabular}
\end{table}

In contrast, the many-to-many models are trained without access to parallel data, taking in input utterances from several speakers (4 speakers, including VCC2SF1 and VCCSM2 in our case, as explained in Section~\ref{sec:dataset}).
This means that the one-to-one and many-to-many models observe very different amounts of data.
Moreover, while the data for both the one-to-one and many-to-many models are divided into a 90\%-10\% train-test split, the same exact splits aren't used in both setups; this is because the former requires parallel utterances, and the split is therefore across utterance \textit{pairs} and not just individual utterances.
To address this, we evaluate the many-to-many models in two settings: on the exact same test utterances as those from the test split of the one-to-one models, as well as on all possible source/target speaker utterance pairs where the source utterance is in the test utterances for the 4 seen training speakers.
In the former case, it could happen that the many-to-many model actually observes one of the test utterances during training.
Nevertheless, reporting scores for both settings allows for a meaningful comparison.

The results of this evaluation on seen speakers are given in Table~\ref{tab:seen-to-seen-results}.
The results indicate that AutoVC appears to be the best in this evaluation on seen speakers.
However, this comes at a computational cost: the linear and StarGAN-ZSVC models are a factor of 5 or more faster than the models relying on recurrent layers like DBLSTM and AutoVC.

\begin{table}[!b]
    \renewcommand{\arraystretch}{1.2}
    \centering
    \caption{Objective evaluation results for zero-shot voice conversion for AutoVC and StarGAN-ZSVC. The \textit{prediction} metrics compare the predicted output to the ground truth target, while the \textit{reconstruction} metrics compare the cyclic reconstruction $X_{\text{src}\rightarrow\text{trg}\rightarrow\text{src}}$ with the original source spectrogram. $e_{\text{norm}}$ indicates the vector norm of the speaker embeddings for the compared spectrograms, with lower values indicating closer speaker identities.}
    \label{tab:zero-shot-results}
    \begin{tabular}{c@{\hspace{0.2cm}}l @{\hspace{0.3cm}} c@{\hspace{0.2cm}}c@{\hspace{0.2cm}}c @{\hspace{0.5cm}} c@{\hspace{0.2cm}}c@{\hspace{0.2cm}}c}
    
    \toprule
    & & \multicolumn{3}{c}{Prediction} & \multicolumn{3}{c}{Reconstruction}\\
    \cmidrule(r){3-5} \cmidrule(r){6-8}
    
    Setting & Model & MAE & $\cos(\theta)$ & $e_{\text{norm}}$ & MAE & $\cos(\theta)$ & $e_{\text{norm}}$ \\
    \midrule
    \multirow{2}{*}[0pt]{Seen-to-unseen} & AutoVC & 1.246 & \textbf{0.982} & 0.742 & 1.178 & 0.976 & 0.392 \\
    & StarGAN-ZSVC & \textbf{1.030} & \textbf{0.982} & \textbf{0.705} & \textbf{0.197} & \textbf{0.997} & \textbf{0.124}\\
    \midrule
    \multirow{2}{*}[0pt]{Unseen-to-seen} & AutoVC & 1.014 & \textbf{0.986} & \textbf{0.328} & 1.201 & 0.975 & \textbf{0.753}\\
    & StarGAN-ZSVC & \textbf{0.974} & 0.985 & 0.380 & \textbf{0.921} & \textbf{0.986} & 0.760 \\
    \midrule
    \multirow{2}{*}[0pt]{Unseen-to-unseen} & AutoVC & 1.238 & 0.981 & 0.746 & 1.340 & 0.968 & 0.827 \\
    & StarGAN-ZSVC & \textbf{1.079} & \textbf{0.982} & \textbf{0.742} & \textbf{0.921} & \textbf{0.986} & \textbf{0.760} \\
    \bottomrule
    \end{tabular}
\end{table}

\subsection{Zero-shot Conversion}
Next we compare StarGAN-ZSVC and AutoVC in zero-shot settings, where either the source, target, or both source and target speaker are unseen during training. 
Many-to-many models can also be used in a cyclic manner; 
we use such cyclic reconstruction as another objective evaluation metric, where we compare how well the original spectrogram is reconstructed when performing this cyclical mapping of converting from the source speaker to the target speaker and back again.
Results for zero-shot conversion are shown in Table~\ref{tab:zero-shot-results}.

The performance for AutoVC and StarGAN-ZSVC are similar on most metrics for the unseen-to-seen case.
But for the seen-to-unseen case and the unseen-to-unseen case (where both the target and source speakers are new) StarGAN-ZSVC achieves both better prediction and reconstruction scores.
This, coupled with its fast inference speed (Section~\ref{sec:experiments_seen}), enables it to be used efficiently and effectively for downstream data augmentation purposes.

\begin{figure}[!t]
\includegraphics[width=\textwidth]{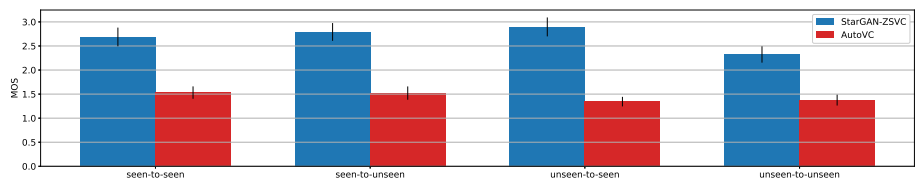}
\caption{Mean opinion score for naturalness for AutoVC and StarGAN-ZSVC in various source/target seen/unseen speaker pairings with 95\% confidence intervals shown.} \label{fig:subjective-eval}
\end{figure}

The results of the subjective evaluation are given in Figure~\ref{fig:subjective-eval}.
To put the values into context, the MOS for the raw source utterances and vocoded source
utterances included in the analysis are 4.86 and 4.33 respectively -- these serve as an upper bound for the MOS values for both models.
Figure~\ref{fig:subjective-eval} largely supports the objective evaluations, providing further evidence that StarGAN-ZSVC outperforms AutoVC in zero-shot settings.
Interestingly, it would appear that StarGAN-ZSVC also appears more natural in the traditional seen-to-seen case.
This evaluation indicates that, for human listeners, StarGAN-ZSVC appears more natural in the low-resource context considered in this paper.

%% file: 6_conclusion.tex
\section{Conclusion}

This paper aimed to improve recent voice conversion methods in terms of speed, the use of non-parallel training data, and zero-shot prediction capability.
To this end, we adapted the existing StarGAN-VC2 system by using a speaker encoder to generate speaker embeddings which are used to condition the generator and discriminator network on the desired source and target speakers.
The resulting model, StarGAN-ZSVC, can perform zero-shot inference and is trainable with non-parallel data.
In a series of experiments comparing StarGAN-ZSVC to the existing zero-shot voice conversion method AutoVC, we demonstrated that StarGAN-ZSVC is at least five times faster than AutoVC, while yielding better scores on objective and subjective metrics in a low-resource zero-shot voice conversion setting. 

For future work, we 
plan to investigate whether scaling StarGAN-ZSVC up to larger datasets yields similar performance to existing high-resource voice conversion
systems, and whether the system could be applied to other tasks aside from pure voice conversion (such as emotion or pronunciation conversion).

%% file: Paper_50.bbl
\begin{thebibliography}{10}
\providecommand{\url}[1]{\texttt{#1}}
\providecommand{\urlprefix}{URL }
\providecommand{\doi}[1]{https://doi.org/#1}

\bibitem{gru}
Cho, K., van Merrienboer, B., Gulcehre, C., Bahdanau, D., Bougares, F.,
  Schwenk, H., Bengio, Y.: {Learning Phrase Representations using RNN
  Encoder–Decoder for Statistical Machine Translation}. EMNLP  (2014)

\bibitem{choi2018stargan}
Choi, Y., Choi, M., Kim, M., Ha, J.W., Kim, S., Choo, J.: {StarGAN: Unified
  Generative Adversarial Networks for Multi-Domain Image-to-Image Translation}.
  In: IEEE CVPR (2018)

\bibitem{vqvae-comparisons}
{Chorowski}, J., {Weiss}, R.J., {Bengio}, S., {van den Oord}, A.: {Unsupervised
  speech representation learning using WaveNet autoencoders}. arXiv e-prints
  arXiv:1901.08810 (2019)

\bibitem{glu}
Dauphin, Y.N., Fan, A., Auli, M., Grangier, D.: {Language Modeling with Gated
  Convolutional Networks}. In: Precup, D., Teh, Y.W. (eds.) PMLR (2017)

\bibitem{CIN}
Dumoulin, V., Shlens, J., Kudlur, M.: A learned representation for artistic
  style. In: {ICLR} (2017)

\bibitem{rules-vc-2}
Erro, D., Moreno, A.: {Weighted frequency warping for voice conversion}. In:
  Interspeech (2007)

\bibitem{xresnet}
He, T., Zhang, Z., Zhang, H., Zhang, Z., Xie, J., Li, M.: {Bag of Tricks for
  Image Classification with Convolutional Neural Networks}. In: IEEE CVPR
  (2019)

\bibitem{lstm}
Hochreiter, S., Schmidhuber, J.: {Long Short-term Memory}. Neural Computation
  \textbf{9},  1735--80 (12 1997)

\bibitem{fastai:unet}
Howard, J., Gugger, S.: {DynamicUnet: create a U-Net from a given architecture}
   (2020), \url{https://docs.fast.ai/vision.models.unet#DynamicUnet}, {Last
  accessed 8 Aug 2020}

\bibitem{disguise}
{Huang}, C., {Lin}, Y.Y., {Lee}, H., {Lee}, L.: {Defending Your Voice:
  Adversarial Attack on Voice Conversion}. arXiv e-prints arXiv:2005.08781
  (2020)

\bibitem{stargan-vc}
Kameoka, H., Kaneko, T., Tanaka, K., Hojo, N.: {StarGAN-{VC}: Non-parallel
  many-to-many voice conversion using star generative adversarial networks}.
  In: IEEE SLT Workshop (2018)

\bibitem{autoencoder-vc2}
Kameoka, H., Kaneko, T., Tanaka, K., Hojo, N.: {ACVAE-VC: Non-Parallel Voice
  Conversion With Auxiliary Classifier Variational Autoencoder}. IEEE
  Transactions on Audio, Speech, and Language Processing  \textbf{27}(9),
  1432–1443 (2019)

\bibitem{stargan-vc2}
Kaneko, T., Kameoka, H., Tanaka, K., Hojo, N.: {{StarGAN-{VC2}: Rethinking
  Conditional Methods for StarGAN-Based Voice Conversion}}. In: Interspeech
  (2019)

\bibitem{adam}
{Kingma}, D.P., {Ba}, J.: {Adam: A Method for Stochastic Optimization}. arXiv
  e-prints arXiv:1412.6980 (2014)

\bibitem{MelGAN}
Kumar, K., Kumar, R., de~Boissiere, T., Gestin, L., Teoh, W.Z., Sotelo, J.,
  de~Br\'{e}bisson, A., Bengio, Y., Courville, A.C.: {MelGAN}: Generative
  adversarial networks for conditional waveform synthesis. In: NeurIPS (2019)

\bibitem{privacy}
{Lal Srivastava}, B.M., {Vauquier}, N., {Sahidullah}, M., {Bellet}, A.,
  {Tommasi}, M., {Vincent}, E.: {Evaluating Voice Conversion-Based Privacy
  Protection against Informed Attackers}. In: ICASSP (2020)

\bibitem{vcc2018}
Lorenzo-Trueba, J., Yamagishi, J., Toda, T., Saito, D., Villavicencio, F.,
  Kinnunen, T., Ling, Z.: {The Voice Conversion Challenge 2018: Promoting
  Development of Parallel and Nonparallel Methods }. In: Odyssey Speaker and
  Language Recognition Workshop (2018)

\bibitem{LSGAN}
Mao, X., Li, Q., Xie, H., Lau, R.Y., Wang, Z., Smolley, S.P.: {Least Squares
  Generative Adversarial Networks}. ICCV  (2017)

\bibitem{projection_discriminator}
Miyato, T., Koyama, M.: {c{GAN}s with Projection Discriminator}. In: ICLR
  (2018)

\bibitem{WORLD-vocoder}
Morise, M., Yokomori, F., Ozawa, K.: {WORLD: A Vocoder-Based High-Quality
  Speech Synthesis System for Real-Time Applications}. IEICE Transactions on
  Information and Systems  \textbf{E99.D}(7),  1877–1884 (2016)

\bibitem{wavenet}
van~den Oord, A., Dieleman, S., Zen, H., Simonyan, K., Vinyals, O., Graves, A.,
  Kalchbrenner, N., Senior, A., Kavukcuoglu, K.: {WaveNet: A Generative Model
  for Raw Audio}. arXiv e-prints arXiv:1609.03499 (2016)

\bibitem{librispeech}
{Panayotov}, V., {Chen}, G., {Povey}, D., {Khudanpur}, S.: {Librispeech: An ASR
  corpus based on public domain audio books}. In: IEEE ICASSP (2015)

\bibitem{waveglow}
{Prenger}, R., {Valle}, R., {Catanzaro}, B.: {Waveglow: A Flow-based Generative
  Network for Speech Synthesis}. In: IEEE ICASSP (2019)

\bibitem{autovc}
Qian, K., Zhang, Y., Chang, S., Yang, X., Hasegawa-Johnson, M.: {A}uto{VC}:
  Zero-shot voice style transfer with only autoencoder loss. In: PMLR (2019)

\bibitem{convoice}
{Rebryk}, Y., {Beliaev}, S.: {ConVoice: Real-Time Zero-Shot Voice Style
  Transfer with Convolutional Network}. arXiv e-prints arXiv:2005.07815 (2020)

\bibitem{unet}
Ronneberger, O., Fischer, P., Brox, T.: {U-Net: Convolutional Networks for
  Biomedical Image Segmentation}. In: MICCAI. Springer International Publishing
  (2015)

\bibitem{rules-vc-1}
Shuang, Z.W., Bakis, R., Shechtman, S., Chazan, D., Qin, Y.: {Frequency warping
  based on mapping formant parameters}. In: Interspeech (2006)

\bibitem{metaanalysis}
{Sisman}, B., {Yamagishi}, J., {King}, S., {Li}, H.: {An Overview of Voice
  Conversion and its Challenges: From Statistical Modeling to Deep Learning}.
  arXiv e-prints arXiv:2008.03648 (2020)

\bibitem{smith:cyclic}
{Smith}, L.N.: {Cyclical Learning Rates for Training Neural Networks}. In: IEEE
  WACV (2017)

\bibitem{gmm-vc-2}
{Stylianou}, Y., {Cappe}, O., {Moulines}, E.: {Continuous probabilistic
  transform for voice conversion}. IEEE Transactions on Speech and Audio
  Processing  \textbf{6}(2),  131--142 (1998)

\bibitem{BDLSTM}
{Sun}, L., {Kang}, S., {Li}, K., {Meng}, H.: {Voice conversion using deep
  Bidirectional Long Short-Term Memory based Recurrent Neural Networks}. In:
  IEEE ICASSP (2015)

\bibitem{vtln-vc}
Suundermann, D., Strecha, G., Bonafonte, A., Höge, H., Ney, H.: Evaluation of
  vtln-based voice conversion for embedded speech synthesis. In: Interspeech
  (2005)

\bibitem{gmm-vc-1}
{Toda}, T., {Black}, A.W., {Tokuda}, K.: {Voice Conversion Based on
  Maximum-Likelihood Estimation of Spectral Parameter Trajectory}. IEEE
  Transactions on Audio, Speech, and Language Processing  \textbf{15}(8),
  2222--2235 (2007)

\bibitem{vctk}
Veaux, C., Yamagishi, J., Macdonald, K.: {CSTR VCTK Corpus: English
  Multi-speaker Corpus for CSTR Voice Cloning Toolkit} (2017),
  \url{http://homepages.inf.ed.ac.uk/jyamagis/page3/page58/page58.html}, {Last
  accessed 1 Sep 2020}

\bibitem{GE2E}
Wan, L., Wang, Q., Papir, A., Moreno, I.L.: {Generalized End-to-End Loss for
  Speaker Verification}. ICASSP  (2018)

\bibitem{vcc2020}
{Zhao}, Y., {Huang}, W.C., {Tian}, X., {Yamagishi}, J., {Das}, R.K.,
  {Kinnunen}, T., {Ling}, Z., {Toda}, T.: {Voice Conversion Challenge 2020:
  Intra-lingual semi-parallel and cross-lingual voice conversion}. arXiv
  e-prints arXiv:2008.12527 (2020)

\bibitem{vc_vc_spk_id}
{Zhizheng}, W., {Haizhou}, L.: {Voice conversion versus speaker verification:
  an overview}. APSIPA Transactions on Signal and Information Processing
  \textbf{3}, ~e17 (2014)

\end{thebibliography}
